\documentclass[twocolumn, twocolappendix]{aastex631}

\usepackage{comment}
\usepackage{hyperref}

\definecolor{indigo}{rgb}{0.0, 0.25, 0.42}
\definecolor{forestgreen}{rgb}{0.13, 0.55, 0.13}
\def\asec{$^{\prime\prime}$~}
\shorttitle{JWST IRAS04302}
\shortauthors{Villenave et al.}
\graphicspath{{./}{figures/}}

\begin{document}

\title{JWST imaging of edge-on protoplanetary disks \\II. Appearance of edge-on disks with a tilted inner region: case study of IRAS04302+2247 }

\email{marion.f.villenave@jpl.nasa.gov}
\author[0000-0002-8962-448X]{Marion Villenave}
\affiliation{Jet Propulsion Laboratory, California Institute of Technology, 4800 Oak Grove Drive, Pasadena, CA 91109, USA}

\author[0000-0002-2805-7338]{Karl R. Stapelfeldt}
\affiliation{Jet Propulsion Laboratory, California Institute of Technology, 4800 Oak Grove Drive, Pasadena, CA 91109, USA}

\author[0000-0002-2805-7338]{Gaspard Duch\^ene}
\affiliation{Astronomy Department, University of California, Berkeley, CA 94720, USA}
\affiliation{Univ. Grenoble Alpes, CNRS, IPAG, F-38000 Grenoble, France}

\author[0000-0002-1637-7393]{Fran\c{c}ois M\'enard}
\affiliation{Univ. Grenoble Alpes, CNRS, IPAG, F-38000 Grenoble, France}

\author[0000-0002-9977-8255]{Schuyler G. Wolff}
\affiliation{Department of Astronomy and Steward Observatory, University of Arizona, Tucson, AZ 85721, USA}

\author[0000-0002-3191-8151]{Marshall D. Perrin}
\affiliation{Space Telescope Science Institute, Baltimore, MD 21218, USA}

\author[0000-0001-5907-5179]{Christophe Pinte}
\affiliation{School of Physics and Astronomy, Monash University, Clayton Vic 3800, Australia}
\affiliation{Univ. Grenoble Alpes, CNRS, IPAG, F-38000 Grenoble, France}

\author[0000-0003-1451-6836]{Ryo Tazaki}
\affiliation{Univ. Grenoble Alpes, CNRS, IPAG, F-38000 Grenoble, France}

\author[0000-0001-5334-5107]{Deborah L. Padgett}
\affiliation{Jet Propulsion Laboratory, California Institute of Technology, 4800 Oak Grove Drive, Pasadena, CA 91109, USA}

\begin{abstract}
We present JWST imaging from 2\,$\mu$m to 21\,$\mu$m of the edge-on protoplanetary disk around the embedded young star IRAS04302+2247. 
The structure of the source shows two reflection nebulae separated by a dark lane. The source extent is dominated by the extended filamentary envelope at $\sim$4.4$\mu$m and shorter wavelengths, transitioning at 7$\mu$m and longer wavelengths to more compact lobes of scattered light from the disk itself.  
The dark lane thickness does not vary significantly with wavelength, which we interpret as an indication for intermediate-sized ($\sim10\mu$m) grains in the upper layers of the disk. 
Intriguingly, we find that the brightest nebula of IRAS40302 switches side between 12.8\,$\mu$m and 21\,$\mu$m. We explore the effect of a tilted inner region on the general appearance of edge-on disks. We find that radiative transfer models of a disk including a tilted inner region can reproduce an inversion in the brightest nebula. In addition, for specific orientations, the model also predicts strong lateral asymmetries, which can occur for more than half possible viewing azimuths. A large number of edge-on protoplanetary disks observed in scattered light show such lateral asymmetries (15/20), which suggests that a large fraction of protoplanetary disks might host a tilted inner region. Stellar spots may also induce lateral asymmetries, which are expected to vary over a significantly shorter timescale. Variability studies of edge-on disks would allow to test the dominant scenario for the origin of these asymmetries.
\end{abstract}


\section{Introduction}

Exoplanets statistics have shown that most stars have a planet orbiting them \citep{Winn_2015, Vigan_2021}. 
Because planet formation occurs in the protoplanetary disk phase, studying protoplanetary disk evolution can allow to better understand planet formation. In particular, it is not fully understood yet how planets can grow from micron-sized dust to kilometer-sized bodies in a few million years, or possibly less. In the current paradigm, high dust concentrations are thought to accelerate grain growth by promoting disk instabilities that lead to planetesimal formation \citep[e.g., streaming instability][]{Youdin_2007}, and subsequently allowing efficient growth via pebble accretion~\citep{Lambrechts_2012}.

Dust vertical settling in the disk is the result of gas drag on dust grains subject to stellar gravity and gas turbulence. This mechanism leads large dust grains to fall into the disk midplane and accumulate there, which is favorable for planet formation. However, this mechanism remains poorly observationally constrained. For relatively large dust particles visible at millimeter wavelengths, it is currently accepted that vertical settling is efficient in the Class II phase, that is typically for disks without a remaining envelope. Although still limited to a small number of systems, the disk emission of Class~II systems at millimeter wavelengths appears geometrically thin, with scale height typically less than 4au at a radius of 100au~\citep{Pinte_2016, Doi_Kataoka_2021, Villenave_2022, Liu_2022, Pizzati_2023}. Yet, no clear observational constraints have been obtained regarding smaller-sized dust particles. In particular, the size of the smallest grains that do experience vertical settling is currently unknown.

Furthermore, limited information is currently available regarding the timescale of vertical settling and its efficiency within younger systems, which still possess part of their primordial envelope. The recent results from the eDISK large program~\citep{Ohashi_2023} and two recent independent studies on the embedded disk IRAS04302+2247~\citep{Villenave_2023, Lin_2023}, indicated that large dust is not very settled in Class 0 and I systems. This suggests that most vertical settling of large ($\sim$ millimeter) sized particles might occur between the Class I and Class II phases. 

Highly inclined protoplanetary disks are favorable targets to investigate this mechanism because they allow a direct view of the disk's vertical structure. 
This paper is part of an ongoing James Webb Space Telescope (JWST) near- to mid-infrared imaging campaign of the largest edge-on disks in nearby star-forming regions (GO programs 2562 and 4290 in Cycles~1 and 2, co-PIs: F. M\'enard and K. R. Stapelfeldt). By looking at edge-on disks at near- to mid-infrared wavelengths, the goal of the program is to study vertical dust settling and grain evolution. 
While the programs mostly include Class~II disks, we focus here on the study of the youngest object of the cycle 1 survey, the embedded IRAS04302+2247 (heareafter IRAS04302).  

The system is located in the L1536 cloud in the Taurus star-forming region \citep[$d=161 \pm 3$ pc,][]{Galli_2019} and orbits around a young 1.6 M$_\odot$ star~\citep{Lin_2023}. IRAS04302 has previously been the focus of a number of studies, using observations from the optical to the centimeter.  
The source has been classified as being in the Class I phase, based on the shape of its spectral energy distribution~\citep[e.g.,][]{Kenyon_1995} or its bolometric temperature~\citep{Ohashi_2023}, and consistently with its morphology in scattered light~\citep[e.g.,][]{Lucas_Roche_1997, Padgett_1999, Wolf_2008}. High angular resolution millimeter continuum observations resolved a brightness asymmetry along the minor axis, with the east side being brighter than the west~\citep{Lin_2023}. Similarly, a blueshifted outflow and millimeter molecular line enhancement have also been observed towards the east of the source~\citep{Podio_2020, vant_Hoff_2020}. This indicates that the disk is slightly offset from edge-on, with the eastern side tilted towards us.  
Finally, previous studies also found that its millimeter emission is not significantly concentrated in the midplane \citep{Villenave_2023, Lin_2023}.

In this paper, we present the first mid-infrared observations of the system, obtained with JWST. The data reduction and image results are discussed in Sect.~\ref{sec:reduction} and Sect.~\ref{sec:results}. 
In Sect.~\ref{sec:model}, we construct a toy model to explore the impact of a tilted inner region on the appearance of edge-on disks  and compare the modeling predictions with our JWST observations of IRAS04302. Then, in Sect.~\ref{sec:discussion}, we discuss the implications of our modeling results for previous scattered light observations of a sample of 20 edge-on disks.  In Sect.~\ref{sec:discussion}, we also compare disk appearance for 3 systems in different evolutionary stages observed with JWST, aiming to obtain constraints on vertical settling and grain properties. Finally, we summarize the results in Sect.~\ref{sec:conclusion}.

\section{Data reduction}
\label{sec:reduction}
We observed IRAS04302 with JWST (GO program 2562) using both NIRCam and MIRI instruments in two consecutive visits starting on 2023 February 20 UT 01:20. The observations used 5 different filters, F200W ($2.0\mu$m), F444W ($4.44\mu$m), F770W ($7.7\mu$m), F1280W ($12.8\mu$m), and F2100W  ($21.0\mu$m). The NIRCam images were obtained simultaneously with exposures of 859s. The MIRI observations were taken successively with exposures of 633s at 7.7$\mu$m, 1672s at 12.8$\mu$m and 329s at 21$\mu$m, respectively.  

For the F200W, F770W, F1280W, and F2100W observations, we obtained the phase 2 pipeline calibrated data from the MAST archive, and we re-ran phase 3, using the JWST pipeline version 1.9.5. In the \texttt{skymatch} step, we used the default skymethod \texttt{`global+match'} but set \texttt{subtract = True}. This method first equalizes sky values between the 4 exposures taken in each filter. It then finds a minimum sky value amongst all input images, and subtracts this background value from each frame. The average background levels were 0.36, 11.60, 46.70, and 272.27 MJy/sr in F200W, F770W, F1280W, and F2100W, respectively. 

In addition, the pipeline-reduced MIRI 12.8$\mu$m and 21$\mu$m images show a gradient in the sky brightness between the top and bottom parts of each frame. Because it appears in each individual, dithered, image, it is likely some instrumental artifact. Thus for those two filters, we performed an additional 2D background subtraction step using the \texttt{Background2D} function in the \texttt{photutils} python package.

\begin{figure*}
    \centering
    \includegraphics[width =\textwidth, trim=0cm 0cm 0cm 0cm,clip ]{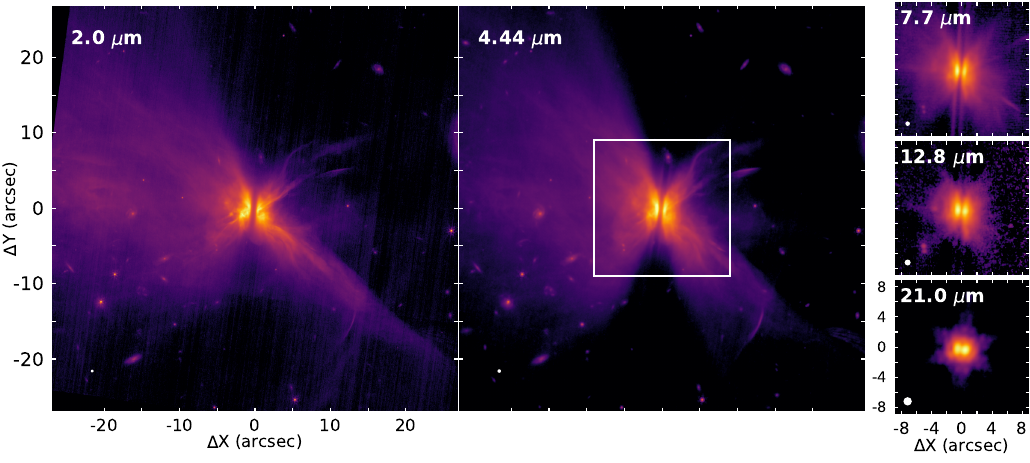}
    \caption{Image gallery of the JWST observations of IRAS04302.  All images are shown with a log stretch. The bottom left ellipse in each panel indicates the corresponding beam size. All images are shown to the same angular scale, and the square box in the 4.44$\mu$m image represent the extent of the 7.7$\mu$m, 12.8$\mu$m, and 21$\mu$m smaller panels. }
    \label{fig:dataIRAS04302}
\end{figure*}

For the F444W observations, we obtained the uncalibrated data from the archive and re-ran each pipeline step using pipeline version  1.11.1.dev16+gb79a88af. In order to recover previously saturated pixels, we set \texttt{suppress\_one\_group = False} in the \texttt{ramp fit} step of phase 1. This allows us to suppress the ramp fit in the case when only the 0$^{th}$ group is unsaturated.  We use the default pipeline values in phase 2 and 3, except for the \texttt{skymatch} step where we subtract the background. The average background level in F444W was 0.49 MJy/sr.

Finally, we resampled and aligned each image to the NIRCAM F200W pixel size and position. We adopt the \texttt{reproject} python package to resample the images, and align them using a background galaxy visible in all fields.  The alignment does not require a rotation, only a translation in right ascension and declination. The final images are presented in Fig.~\ref{fig:dataIRAS04302}. \\

In this work, we also use archival HST NICMOS images, from program 7418 (PI: D. Padgett). The HST observations were obtained using filter F205W ($2.05\mu$m). They were previously published by \citet{Padgett_1999}. These observations were taken on August 19, 1997, nearly 25.5 years prior to the JWST data.  For this study, we use the data directly downloaded from the archive.

\section{Results}
\label{sec:results}
\subsection{Disk morphology }
\label{sec:morphology}

\begin{table*}
    \caption{IRAS04302 morphological properties.}
    \centering
    \begin{tabular}{lccccccccccc}
    \hline\hline
    $\lambda$ & $F_{20^{\prime\prime} \times 20^{\prime\prime}}$ &$d_{neb}$ & FR &  $R_{FWHM}$ &$R_{FWHM}$ & $R_{FW10\%}$ & $R_{FW10\%}$ & $\delta_{spines}$\\ 
    ($\mu$m)& (mJy)& (arcsec) & W/E& E  (arcsec) & W  (arcsec) &E  (arcsec) & W (arcsec) & W-E (\%)\\
    \hline
    2.00$^{(E)}$ & $17 \pm 2$ & $  1.35 \pm 0.10 $ & $ 0.99 \pm 0.01$ & $ 2.17 \pm 0.25 $ & $ 1.75 \pm 0.43 $ &  $ 3.69 \pm 0.13 $ & $ 3.83 \pm 0.17 $ & 12\\
    4.44$^{(E)}$ &  $32 \pm 3$ & $1.13 \pm 0.10$  & $0.58 \pm 0.01$  & $ 1.71 \pm 0.10$  & $1.63 \pm 0.15$  & $2.97 \pm 0.10$  & $3.78 \pm 0.19$ & 11 \\
    \hline
     7.70 & $16 \pm 2$ &$ 1.03 \pm 0.10$  & $0.52 \pm 0.01$  & $ 1.35 \pm 0.14$  & $ 1.63 \pm 0.10$  & $ 2.76 \pm 0.10$  & $ 3.57 \pm 0.10$ & 7\\
     12.8 & $11 \pm 1$ &$1.10 \pm 0.10$  & $ 0.78 \pm 0.04$  & $ 1.30 \pm 0.16$  & $ 1.43 \pm 0.10$  & $2.69 \pm 0.10$  & $3.12 \pm 0.15$ & 8\\
     21.0& $101 \pm 10$ &$ 0.96 \pm 0.10$  & $ 1.34 \pm 0.05 $  & $1.31 \pm 0.10$  & $ 1.17 \pm 0.10$  & $-$  & $-$ &  $-$\\
    \hline
    \end{tabular}
    \tablecomments{$F_{20\arcsec\times20\arcsec}$ are the flux densities integrated over a $20\arcsec\times20\arcsec$ aperture, $d_{neb}$ the darklane width, FR the peak flux density ratio of the west over the east nebula, while $R_{FWHM}$ and $R_{FW10\%}$ correspond to the radial size of each nebula, respectively measured at 50\% or 10\% of the peak. $\delta_{spines}$ characterizes the difference in lateral centering between both nebulae. 
    At 21$\mu$m, the signal to noise is not sufficient to measure $R_{FW10\%}$.  
    The error bars of $d_{neb}$, $R_{FWHM}$, and $R_{FW10\%}$ correspond to the statistical error for the different spine averaging or 0.1\asec, whichever is greater.
    A positive value of $\delta_{spines}$ indicates that the west nebula is more centered towards the south than the east nebula.  $^{(E)}$ indicates that the images are dominated by the envelope; at these wavelengths our measurement might not trace the disk properties.}
    \label{tab:darklane}
\end{table*}

We show the 5 new JWST images of IRAS04302 in Fig.~\ref{fig:dataIRAS04302}. All images present a similar overall shape, displaying two extended nebulae, separated by a dark lane. This morphology is typical of scattered light observations of edge-on sources, indicating that scattering dominates up to 21$\mu$m. However, we note that the origin of the photons likely varies between the different wavelengths. Indeed, while at short wavelengths (e.g., 2$\mu$m) the star dominates, thermal emission from the hot inner disk progressively becomes important over the stellar emission as the wavelength increases towards the mid-infrared. 
The morphology of the images show that the disk is optically thick, even to its own radiation, up to 21$\mu$m.

Besides a globally similar overall shape, the images reveal a clear change of morphology around 7.7$\mu$m, where the apparent size of the source gets dramatically smaller. This marks the limit up to which scattering off the envelope dominates over scattering off the disk. At short wavelengths ($2.0\mu$m, $4.44\mu$m), the scattered light reveals two nebulae surrounded by diffuse extended emission which includes some filamentary features. At 2.0$\mu$m, both the east nebula and the south west filament are visible at distances greater than 20\asec from the central dark lane. 

At 7.7$\mu$m, the source appears much more compact, with most diffuse emission from shorter wavelengths observations not detectable. Two nearly vertical lines, along the major axis direction, are also visible at this wavelength. We note however that these lines are not physical, but instead they are related to the known `cross artifact' at this wavelength~\citep[][also refered to as the cruciform artifact in the MIRI user documentation]{Gaspar_2021}.  

Finally, at longer wavelengths ($12.8\mu$m, 21$\mu$m) the diffuse emission disappears totally below the rising thermal background. At these wavelengths, IRAS04302 shows two smooth nebulae separated by a dark lane. The absence of diffuse emission and smooth shape of the two nebulae suggest that, at these wavelengths, scattering from the disk dominates. In addition, we note that the disk is more centrally peaked and radially smaller at 21$\mu$m than at 12.8$\mu$m, which could be related to dust properties.

For all 5 filters, we use a 20\asec square aperture to measure the overall source flux density.  We report the results in the first column of Table~\ref{tab:darklane}. 
Then, we follow the  methodology recently highlighted by \citet{Duchene_2023} to measure structural source's parameters~\citep[darklane width, brightness ratio, size, lateral asymmetries, see also Appendix~D of][]{Villenave_2020}.  We parametrize the vertical position of the two scattering surfaces as a function of radius at each wavelength. This is done up to a radius of 1\asec at 21$\mu$m, or 3\farcs4 at the other wavelengths, and up to 0.8\asec vertically from the midplane for all images.  
To do so, we take minor axis profiles averaged over 0.4\asec or 0.5\asec along the major axis of the source.
For each nebula, we determine the position of the peak by fitting a polynomial of degree 6 or~8.   
This allows us to obtain four sets of arrays with the peaks coordinates of the east and west nebulae, to which we fit a second order polynomial to generate the final spines. 

We use these polynomial spines to estimate the  dark lane thickness (closest distance between the nebulae, $d_{neb}$), 
the peak flux ratio of the west over the east nebula (FR), radial full width half maximum ($R_{FWHM}$) and the radial full width at 10\% of the peak ($R_{FW10\%}$)  of each spine. In addition, we characterize the lateral offset between the spines by the parameter $\delta_{spines}$. This parameter quantifies the difference between the median location of each spine, within 10\% of their respective peaks. To allow comparison with disks of different sizes, the lateral difference is normalized by the averaged disk size (i.e., mean $R_{FW10\%}$, respectively at each wavelength). This final number is reported as $\delta_{spines}$.
 
We present the average of these values obtained with the four sets of spines and corresponding standard deviation in Table~\ref{tab:darklane}. We note that when building the spine without the radial averaging and polynomial fitting $d_{neb}$ is consistently found to be $\sim0\farcs1$ ($\sim$ 3 pixels) smaller than the values reported in Table~\ref{tab:darklane}, except at 2.0$\mu$m where it is $0\farcs3$ broader (highlighting the uncertainty of the methodology when the image appears more messy). Moreover, we caution that for the shortest wavelengths,  $\lambda <7\mu$m, the images are likely dominated by the envelope such that our measurements might not trace the disk properties.

\begin{figure*}
    \centering
    \includegraphics[width = 0.95\textwidth, trim=0.7cm 0.1cm 0cm 0.4cm,clip ]{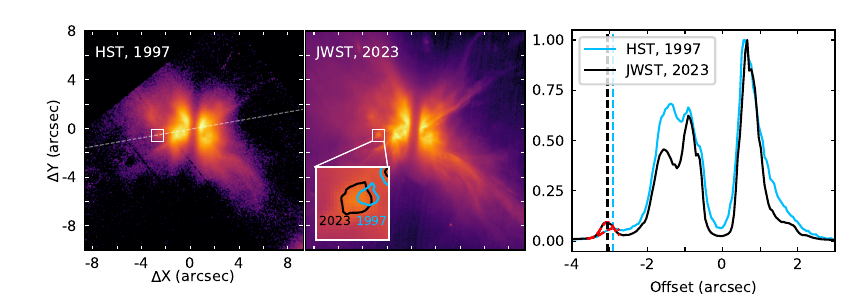}
    \caption{Comparison of the 2.05$\mu$m HST (left panel) and 2.0$\mu$m unconvolved JWST (middle panel), previously aligned, images of IRAS04302. The contours correspond to intensity level of 5\% of the peak (blue is HST, black is JWST), and the white line in the left panel shows the orientation of the normalized cut, displayed in the right panel. A 1D Gaussian was fitted to both cuts (marked in red), and the vertical lines in the right panel indicate the resulting centering of these gaussians for each dataset.}
    \label{fig:blob}
\end{figure*}

Focusing on the wavelengths $>7\mu$m which are mainly sensitive to the disk, Table~\ref{tab:darklane} shows that the west side of the disk is more radially extended  than the east side, both at 7.7$\mu$m and 12.8$\mu$m. In addition, the value of $\delta_{spines}$ is of order 10\%, which illustrates that the two nebulae are not perfectly aligned with each other. The west nebula is found to extend further south that the east nebula at 7.7$\mu$m and 12.8$\mu$m. Lastly, we find that the dark lane thickness does not vary significantly with wavelength (by less than 10\% for $\lambda >4\mu$m), which we discuss further in Sect.~\ref{sec:comparison_darklane}.

Surprisingly, we also notice that the brightest nebula flips with wavelength. At 7.7$\mu$m and 12.8$\mu$m, the east nebula is brighter than the west nebula, while the opposite is true at 21$\mu$m, where the west nebula is the brightest. We note that Table~\ref{tab:darklane} reports the peak brightness ratio, but this is also true for the integrated brightness of both nebulae. 
In Sect.~\ref{sec:model} we perform radiative transfer models aiming to reproduce this effect, and we discuss the results in Sect.~\ref{sec:discussion_tilt}.

\subsection{Comparison of two epochs}

IRAS04302 was observed multiple times at $\sim2\mu$m \citep{Lucas_Roche_1997, Padgett_1999}. In this section, we compare the 2023 JWST observations and the previous 1997 HST 2.05$\mu$m image of \citet{Padgett_1999}. We aim to look for moving features between the two observations. 

To do so, we first aligned the two images. Because no stars were visible in the background of the HST image, we cross-correlated the HST and JWST images to align them. We first resampled the JWST NIRCAM image to the coarser HST NICMOS pixel scale. We normalized the JWST and HST images to their respective total counts, and, to align the images, we then shifted the JWST data to minimize the normalized residuals. We perform this alignement both before and after convolving the JWST observations with a HST PSF, generated using the TinyTim software~\citep{Krist_2011}.

We display the aligned images and significant contours in Fig.~\ref{fig:blob}. The JWST observations reach a signal to noise significantly better than the previous data, such that diffuse emission is seen to a much larger distance. Filaments features that were barely visible with HST are now confirmed with the new image. In addition, we observe some differences in the source's morphology. In particular, around -1.5'' eastward of the star, the  source is dimmer in the JWST observations than in the older HST images. 
This is possibly due to time variation of the density distribution in the nebula and/or variable illumination from the  star~\citep[e.g.,][]{Watson_2007}.

In addition, we find that a roundish blob towards the east side of the source (see insert in Fig.~\ref{fig:blob}) seems to move towards the outer regions of the source. 
To quantify the movement of this blob feature, we estimate the location of the star in the disk-dominated 21$\mu$m observations, and obtain the fluxes along a cut intercepting this star location and the globular feature. We show the fluxes obtained along this cut, normalized to their respective maximum, in the right panel of Fig.~\ref{fig:blob}. The cuts show two bright peaks with maximum within $\pm$1\arcsec, and a small increase of emission, around -3\arcsec, which corresponds to the globular feature. For each cut, we fitted this small increase by a 1D Gaussian. On the JWST data, this was done twice, both for the HST-convolved image and that with the native resolution. 
We find that the center of this blob feature moves by 0.10\asec (resp. 0.16\arcsec)\footnote{The difference in distance between the convolved and unconvolved data corresponds to $\sim$1 HST pixel. It mostly due to the non-perfectly gaussian shape of the excess.} 
from the star between the HST image and JWST convolved data (resp. native JWST data). This motion is pointing directly away from the presumed location of the star.

The relative proper motion of this blob compared to the disk location is +3.9~mas/yr in right ascension (resp. +6.2 mas/yr), and -0.7 mas/yr in declination (resp. -1.1 mas/yr). Using Gaia and VLBI astrometry, \citet{Galli_2019} estimated that the proper motion of the L1536 cloud in which IRAS04302 is located is +10 mas/yr in right ascension and -17 mas/yr in declination. The proper motion of the blob feature appears unrelated to the proper motion of the disk, which suggests that it is not a background object. 

Assuming a distance to the disk of 161~pc~\citep{Galli_2019}, we estimate that the blob's apparent movement is 16~au (resp. 26~au) in $\sim$ 25.5 years. Thus, we infer an apparent velocity of 3.0~km/s (resp. 4.8 km/s) for this feature. This velocity is significantly less than typical velocities of winds or jets~\citep[e.g.,][]{Pascucci_2022}. 
Yet, this could correspond to a knot of dust entrained in the low-velocity outflow. Indeed, if the dust lies at the outflow cavity edge, its actual velocity could be slightly larger, due to projection effects.

\section{Radiative transfer modeling}
\label{sec:model}

\subsection{Motivations}
\label{sec:motivation}

The new JWST observations of IRAS04302 show that the flux ratio between the two nebulae is not constant with wavelength (Sect.~\ref{sec:morphology}), and in particular, we found that the brightest nebula switches side between 12.8$\mu$m and 21$\mu$m.  This was an unexpected result. Indeed, at wavelengths between the optical and infrared, we expect forward scattering to be important. This implies that a disk that is slightly offset from edge-on will have one side that is brighter than the other, simply due to its orientation, and independent of the wavelength. 
Because the MIRI observations of IRAS04302 were performed consecutively within less than one hour, variability likely does not explain this behavior.

Multi-wavelengths optical and near-infrared observations have previously been performed in about a dozen edge-on disks, mainly at wavelengths between 0.4$\mu$m and 2$\mu$m. As expected in the case of axi-symmetric radiative transfer models, the  vast majority of systems \citep[e.g., Haro6-5B, LkHa 263C, HV Tau C, PDS144N, HH30,  HOPS136, ESO-H$\alpha$569, Oph163131, Tau042021;][]{Krist_1998, chauvin_2002,   Stapelfeldt_2003, Perrin_2006, Watson_2007_HH30, Fischer_2014, Wolff_2017, Wolff_2021, Duchene_2023} do not show a switch in the brightest nebula.

While the nebulae flux ratio in the edge-on disk HK~Tau~B varies significantly between 0.6$\mu$m and 3.8$\mu$m~\citep[by a factor $\sim8$,][]{McCabe_2011}, to our knowledge, the only other edge-on disk where the brightest nebula switches side between two wavelengths is 2MASS J16281370-2431391 (Flying Saucer), although at different wavelengths than in IRAS04302. Using VLT/ISAAC, \citet{Grosso_2003} indeed found that the flux ratio of the two nebulae reverses between 1.3$\mu$m and 2.2$\mu$m. They suggested that a disk warp might be able to explain the differences by introducing some extra self-absorption in the disk. 

\citet{Nealon_2019} performed radiative transfer modeling of numerical simulations including a warp, allowing some initial testing of this hypothesis. They showed that, when a warp is included,  a true edge-on disk ($i=90^\circ$) observed at 1.6$\mu$m can appear asymmetric, with the brightest nebula depending on the angle of observation of the system (their Figure 6).  
Here, we aim to extend this study to multiple wavelengths to analyse the impact of a tilted region in the nebulae flux ratio and other morphological properties in edge-on disks.

\subsection{Model description}

We construct a toy model using the radiative transfer code \texttt{mcfost} \citep{Pinte_2006, Pinte_2009}. 
The goal of this modeling is to illustrate the effect of a tilted inner disk onto edge-on disk appearance. While we construct our model with some disk and stellar parameters consistent with IRAS04302, we do not aim to reproduce this disk. Instead, we are interested in more general disk properties.

We take a stellar effective temperature and radius of $T_{eff} = 4500$K,  $R_\star=3.7$R$_\odot$ ($\sim5L_\odot$), respectively \citep{Grafe_2013, Villenave_2023}. We assume that the disk is broken into two regions. The inner region extends from 0.1 to 5au, while the outer region extends from 9 to 300au. We set the dust masses to  $1.6\ 10^{-6}$~M$_\odot$ and  $5\ 10^{-4}$~M$_\odot$, respectively for the inner and outer disks. 

For both regions, the surface density follows a simple power-law distribution, with $\Sigma(r)\propto r^{-1}$, and the vertical extent is parameterized as $H(r) = 41 (r/300\text{au})^{1.14}$.  
We assume that dust grains are composed of a mixture of 62.5\% of astronomical silicates and 37.5\% of graphite \citep[following][]{Grafe_2013, Villenave_2023},  
and only include small dust grains which follow a power-law distribution $n(a)da\propto a^{-3.5}da$.  We fixed dust sizes between 5 and 30$\mu$m rather that using a broader distribution. 
This choice aims to mimic the amount of forward scattering observed in IRAS04302, as the shape of the brighter side is defined by the forward scattering off the closer surface. We checked that changing the grains properties does not qualitatively alter the discussion below.  Further exploration of the grain properties, including porosity, and aggregate shape will be performed in future work (Tazaki et al. in prep).

\begin{figure}
    \centering
    \includegraphics[width =0.41\textwidth, trim=0cm 0cm 0cm 0cm,clip ]{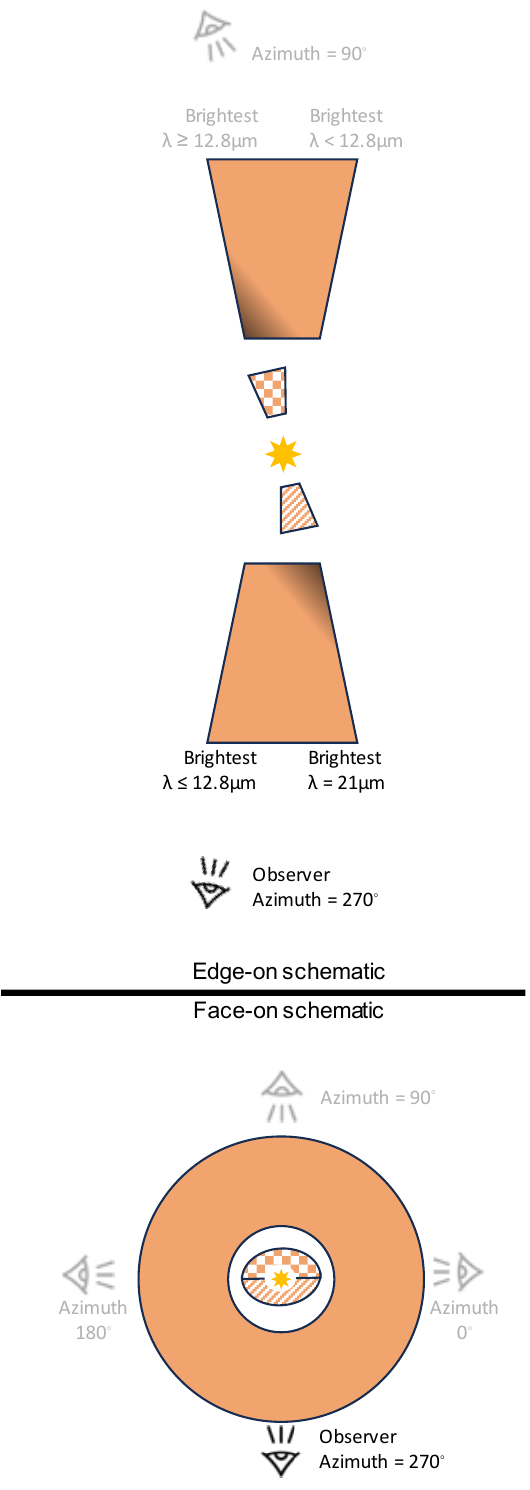}
    \caption{Schematic geometry of the model, not to scale. The orientation corresponding to an azimuth of 270$^\circ$ allows to reproduce the flux reversal between 12.8$\mu$m and 21$\mu$m observed in IRAS04302 data~(see Fig.~\ref{fig:model}). In the bottom view, the observer is slightly above the outer disk midplane, the dashed inner region is below that midplane and the squared inner region is above it. }
    \label{fig:schematic}
\end{figure}

We produce two models, one where the inner region is aligned to the outer disk (``no tilt" or ``axisymmetric" model) and another where the inner region is tilted by 10$^\circ$ compared to the outer disk (``tilted" model). We fix the outer disk's inclination to 88$^\circ$~\citep{Villenave_2023, Lin_2023}, and consider 16 different angles of observations (azimuths) of the system  with the tilted inner disk, separated by $22.5^\circ$.  For different azimuths, the orientation of the outer disk remains the same while the orientation of the tilted inner disk varies. 
Azimuths 0$^\circ$ and 180$^\circ$ correspond to the orientation where the line of nodes defined by the inner disk and outer disk planes is pointed directly at the observer.  A schematic view of the system highlighting four different viewing orientations is shown in Fig.~\ref{fig:schematic}. 

\begin{figure*}
    \centering
    \includegraphics[width =0.47\textwidth, trim=0.4cm 0cm 0cm 0cm,clip ]{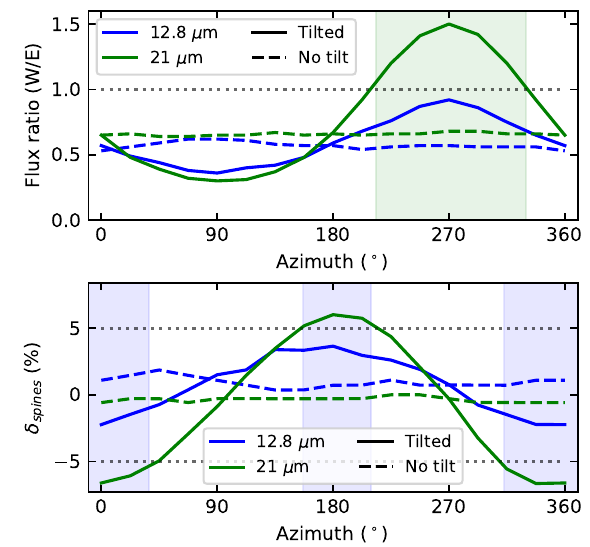}
    \includegraphics[width =0.48\textwidth, trim=0.2cm 0cm 0cm 0cm,clip ]{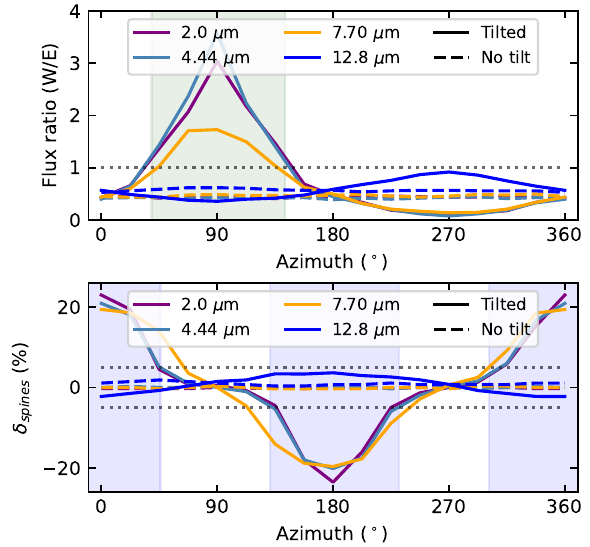} 
    \caption{Variation of flux ratio (top panels) and lateral asymmetries (bottom panels) with azimuth, for the tilted (solid lines) and axisymmetric  (dashed lines) disk models. On the top panels, flux reversal occurs above the horizontal dotted line, and the green region highlights the azimuth where flux is reversed (at 21$\mu$m, left, and at 2$\mu$m, right). The small azimuthal variations ($<15\%$) in the no tilt models are due to  radiative transfer noise. On the bottom panels, the horizontal dotted lines show a value of $\delta_{spines}=\pm5\%$, corresponding to strong lateral asymmetries. The blue regions show the azimuths where this threshold is surpassed (at 21$\mu$m, left, and at 2$\mu$m,  right). The 12.8 micron model is shown on both columns, to scale, for comparison.}
    \label{fig:azimuth}
\end{figure*}

For each model, we produce images at 2.0$\mu$m, 4.44$\mu$m, and 7.7$\mu$m, 12.8$\mu$m, and 21$\mu$m and convolve them by the JWST PSF obtained with the WebbPSF software~\citep{Perrin_2014}. As for the data~(Sect.~\ref{sec:morphology}), the models are dominated by scattering at all wavelengths. We then employ the methodology of Sect.~\ref{sec:morphology} to fit the spines of the models, and measure the variation of the flux ratio and lateral asymmetry $\delta_{spines}$ with the angle of observation (azimuth). We note that for the models, the radial averaging and polynomial fitting did not improve the quality of the spine, so we did not use them. 

\subsection{Results}
\label{sec:modelresults}

In Fig.~\ref{fig:azimuth}, we show how the flux ratio and the lateral asymmetry vary with azimuth, for wavelengths between 2$\mu$m and 21$\mu$m. 
In this section, we first describe the model features for both the dependence of the flux ratio and the lateral asymmetry with azimuth. 
Then, in a second phase, we describe the reasons why a tilted inner disk may lead to such variations.
Throughout this section the wording ``flux reversal" refers to an azimuth where the brightest nebula in the tilted model is opposite to that of the no tilt model. We use ``brightest nebula switch" when the brightest nebula switches side between wavelengths. 

\paragraph{Flux ratio} In the top panels of Fig.~\ref{fig:azimuth}, we find that while the models with no tilt do not show any clear flux ratio change with azimuth (less than 15\% variations, due to radiative transfer noise), the tilted models do show significant variation with the observing azimuth, at all wavelengths.   
In the case where there is no tilt, the brightest nebula of this model is on the east, implying a flux ratio $<$1. While this is also true for most orientations in the tilted models, some azimuthal range show a flux reversal  (flux ratio $>$1) at 21$\mu$m and for $\lambda<12.8\mu$m. No flux reversal is found for the models at 12.8$\mu$m, although there is some enhancement of the flux ratio within the range where the flux reversal is observed at 21$\mu$m.  

At 21$\mu$m, the flux reversal occurs for azimuths between $\sim210-330^\circ$ (see schematic orientation in Fig.~\ref{fig:schematic}), while it is observed for the opposite orientation between 2$\mu$m and 7.7$\mu$m (azimuths between $\sim40-140^\circ$). 
In both cases, the flux reversal only occurs about $\sim$1/3 of the possible viewing orientations.

\paragraph{Lateral asymmetries} In addition to generating flux ratio variations with azimuth, introducing a tilted inner region also leads to lateral asymmetries in the disk. The bottom panels of Fig.~\ref{fig:azimuth} show the azimuthal variation of the parameter $\delta_{spines}$. This parameter, introduced in Sect.~\ref{sec:morphology}, quantifies the difference in centering between the spines of both nebulae, where the spines are only considered up to 10\% of their respective peaks. By visually inspecting the models, we identify that a value of $\delta_{spines}\gtrsim5\%$ leads to significant lateral asymmetry, which can be directly detectable  by eye. 

We find that this threshold is reached for a similar range of orientations at 21$\mu$m and between 2$\mu$m and 7.7$\mu$m. No clear azimuthal variation is found at 12.8$\mu$m.  At 21$\mu$m, this threshold is reached for azimuths between $\sim0-40^\circ$, $160-210^\circ$, $310-360^\circ$. On the other hand, at shorter wavelengths the lateral asymmetries are stronger and overtake the threshold for a wider range of orientations ($\sim0-45^\circ$, $130-230^\circ$, $300-360^\circ$). 
The azimuths where lateral asymmetries are observed in the disk correspond to orientations where the observer is close to perpendicular to the tilted inner region (Fig.~\ref{fig:schematic}).

Interestingly, we find that, at short wavelengths, important lateral asymmetries occur for a larger range of azimuths than the range where flux reversal were obtained: about 200$^\circ$ vs 100$^\circ$. More than 1/2 of the possible orientations present strong lateral asymmetries at 2$\mu$m, while about 1/3 may lead to flux reversal. At short wavelengths, lateral asymmetries can thus be an easier and more frequent tracer of inner disk misalignment than strong flux variations. We also checked that at smaller inclinations (80 \& 85$^\circ$) this statement remains valid, with lateral asymmetries remaining clearer.

\begin{figure}
    \centering
    \includegraphics[width =0.47\textwidth, trim=0.cm 0cm 0cm 0.cm, clip ]{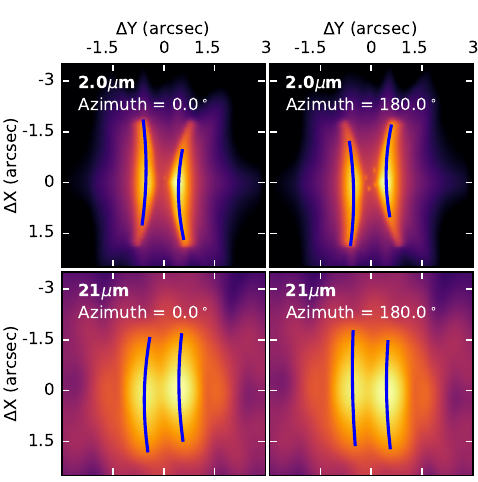}
    \caption{2.0$\mu$m and 21$\mu$m images of the tilted models, for azimuths of 0$^\circ$ and 180$^\circ$. The location of the spines within 10\% of their respective peaks are shown in blue, highlighting clear lateral asymmetries.}
    \label{fig:lateral_asymmetry}
\end{figure}

In Fig.~\ref{fig:lateral_asymmetry}, we illustrate the lateral asymmetries by showing models at 2$\mu$m and 21$\mu$m observed at azimuth 0$^\circ$ and 180$^\circ$, found to be associated to large values of $\delta_{spines}$. The spines, within 10\% of their respective peaks, are represented on top of the models. From this figure, it is clear that the spines are not perfectly centered on each other and that the lateral asymmetry is stronger at shorter wavelengths. Moreover, we also see that the asymmetry is opposite at 2$\mu$m and at 21$\mu$m. For azimuth 0$^\circ$, at 2$\mu$m the east nebula is found to the north of the bottom nebula, while this is the opposite at 21$\mu$m. 
This illustrates that the cause of the asymmetry, which is the same as for the flux reversal, is not the same at 2$\mu$m and 21$\mu$m.

\paragraph{Origins of the azimuthal variations} 
First, we remind that because forward scattering is important at infrared wavelengths, a disk that is slightly offset from edge-on will have one side that is brighter than the other, simply due to its orientation. By definition, this is the east side in our models.

\begin{figure}
    \centering
    \includegraphics[width =0.47\textwidth, trim=0.cm 0cm 0cm 0cm,clip ]{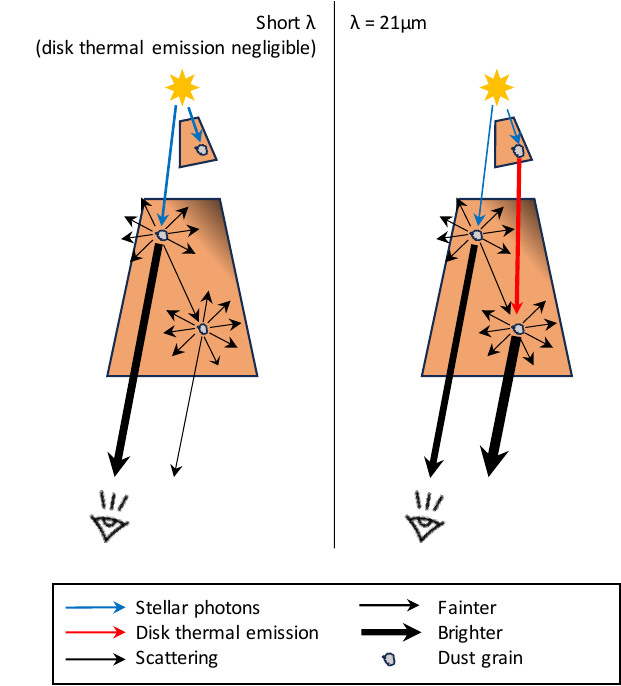}
    \caption{Schematic view of the photon propagation in the front half a disk for two regimes of wavelengths. The observing azimuth corresponds to 270$^\circ$, and can reproduce a flux reversal between 12.8$\mu$m and 21$\mu$m. Forward scattering is more efficient than scattering in other directions as illustrated by the larger arrows.}
    \label{fig:sketch_photonpropagation} 
\end{figure}

At short wavelengths (2.0$\mu$m, 4.44$\mu$m, 7.7$\mu$m), stellar emission and very compact hot disk thermal emission are the dominant source of photons in the disk.  
The tilted inner region however affects the propagation of stellar light to the outer disk. 
In particular, shadows are expected where the inner disk blocks direct starlight (shadowed regions in top panel of Fig.~\ref{fig:schematic}). 
At the same time, the opposite side, that has a direct view to the star, will have more incoming photons and thus it brightness will be enhanced, relatively. For azimuth 270$^\circ$ as pictured in Fig.~\ref{fig:schematic} and Fig.~\ref{fig:sketch_photonpropagation}, this implies that the east nebula is even brighter than for a disk without a tilt. However, for the opposite orientation where the observer stands on top of Fig.~\ref{fig:schematic} (azimuth $\sim90^\circ$), the east side will be in the shadow and thus less bright than the west side. This causes the flux reversal observed in the models. For intermediate orientations (azimuth $\sim0^\circ$ or $\sim180^\circ$, Fig.~\ref{fig:schematic}), the similar effect is seen from the side. The sides in the shadow of the inner disk will appear less radially extended than their counterpart with direct view to the star.

\begin{figure*}
    \centering
    \includegraphics[width = 0.75\textwidth, trim=0.cm 0cm 0cm 0cm, clip ]{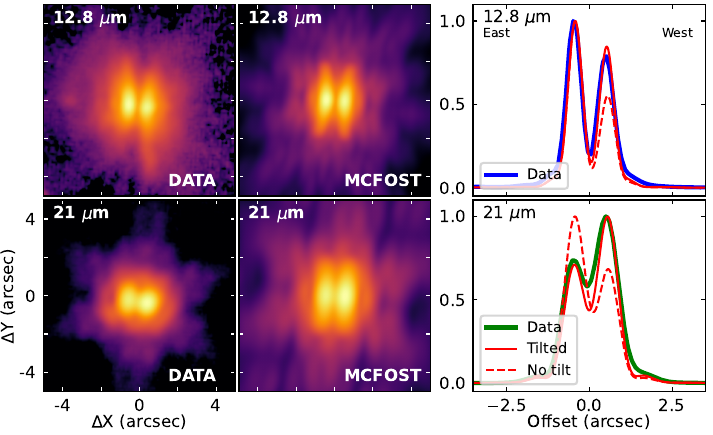}
    \caption{\emph{Left panels:}  JWST 12.8$\mu$m and 21$\mu$m observations of IRAS04302. \emph{Middle panels:} Tilted disk model at 12.8$\mu$m and 21$\mu$m, observed with an azimuth of 270$^\circ$. \emph{Right panels:} Minor axis cut at the center of the disk for the data, tilted, and no tilt models, at 12.8$\mu$m and 21$\mu$m.} 
    \label{fig:model}
\end{figure*}

At longer wavelengths, however, the situation is different because thermal emission from the inner disk dominates over the stellar emission and originates from a radially extended region. 
In fact, the ratio of scattered photons originating from inner disk thermal emission to those originating from the star is more that 2.5 times higher at 21$\mu$m than at 12.8$\mu$m  (ratios of 4300 vs 1600).
The inner disk thermal emission, which is spatially very different from the central star, is added to stellar light, and scattered through the outer disk.  Thus, the outer disk's side towards which the small inner disk is pointing will receive $-$ and thus re-scatter $-$ significantly more photons than the other side and appear brighter.  
For an azimuth of 270$^\circ$ as pictured in Fig.~\ref{fig:schematic}  and Fig.~\ref{fig:sketch_photonpropagation}, this implies that the west nebula is brighter than the east nebula and causes the observed flux reversal. At the opposite orientation (azimuth $\sim90^\circ$), the tilt enhances the east side. At intermediate orientations, we expect lateral asymmetries, but on the opposite side as for short wavelengths. 
In our models, we find that this effect is more important at 21$\mu$m than at 12.8$\mu$m because thermal emission is more important.\\

In this section, we have demonstrated that a tilted inner region can have a strong effect on the appearance of disks viewed close to edge-on at near- to mid-infrared wavelengths. On the one hand, the tilted inner region can lead to a change of the brightest nebula for about 1/3 of the possible disk orientation. Such flux reversal occurs for opposite orientation in the mid-infrared, where thermal emission from the inner disk is important, than in the near-infrared, where stellar photons dominate. 
Moreover, a tilted inner region can lead to strong lateral asymmetries. This effect is stronger at near-infrared wavelengths and occurs for more than 1/2 of the possible viewing orientations. Lastly, lateral asymmetries occur over a similar orientation range at near- and mid-infrared, but have opposite directions.  
While, quantitatively, the presence/degree of reversal and amplitude of lateral asymmetry are dependent on model parameters (e.g., dust properties, tilted inner disk extent and misinclination, viewing geometry), we note that the qualitative behavior is quite general for small misalignements, likely as long as they cast shadows over a broad range of azimuthal angles~\citep[e.g., HD139614,][]{Muro-Arena_2020}.

\subsection{Application to IRAS04302}

\label{sec:caseIRAS04302}
The observations of IRAS04302 show a clear switch in brightest nebula between 12.8$\mu$m and 21$\mu$m (Sect.~\ref{sec:morphology}). In Sect.~\ref{sec:modelresults}, we demonstrated that the presence of a tilted inner region can produce such a brightest nebula switch. Now, in Fig.~\ref{fig:model}, we compare the data with the tilted and no tilt models observed at an azimuth of 270$^\circ$ (see Fig.~\ref{fig:schematic}). We find that the tilted model reproduces the switch with a similar amplitude than the data. This would imply that the east side is tilted slightly toward us, and closest to the observer.

This proposed orientation is consistent with the conclusions from \citet{Lin_2023}. Indeed, using high angular resolution 1.3mm observations, they resolved an asymmetry in the continuum emission of IRAS04302, finding that the east side is brighter than the west side. They concluded that the east side traces the backside of the disk at millimeter wavelengths.  
Besides, the blueshifted outflow and millimeter molecular line enhancement have been observed towards the east in a number of previous studies~\citep{Podio_2020, vant_Hoff_2020, Lin_2023}. These observations are consistent with a moderate tilt and east side of the outer disk tilted towards us.\\

We note however that our toy model does not match every observational feature. On the one hand, the modeled emission is too radially extended at 21$\mu$m compared to the data, suggesting that forward scattering is more important in the data than in the model. Future modeling will be required to infer the dust properties of these systems (Tazaki et al. in prep).

On the other hand, while we show that the current tilted model observed at an azimuth of 270$^\circ$ is able to reproduce the brightest nebula switch between 12.8$\mu$m and 21$\mu$m, our analysis in Sect.~\ref{sec:morphology} also found that IRAS04302 has some degree of lateral asymmetry. This suggests that the disk is not viewed perfectly from an azimuth of 270$^\circ$ but instead with some viewing orientation between 270 and 360$^\circ$.

Finally, it is important to remind that at wavelengths shorter than $\sim7.7\mu$m the envelope is important in IRAS04302. No envelope is included in our models, which can thus not be directly compared with the observations of this system at short wavelengths.

\section{Discussion}
\label{sec:discussion}

\subsection{Tilted inner regions in protoplanetary disks}
\label{sec:discussion_tilt}

\begin{figure*}
    \centering
    \includegraphics[width = 0.9\textwidth]{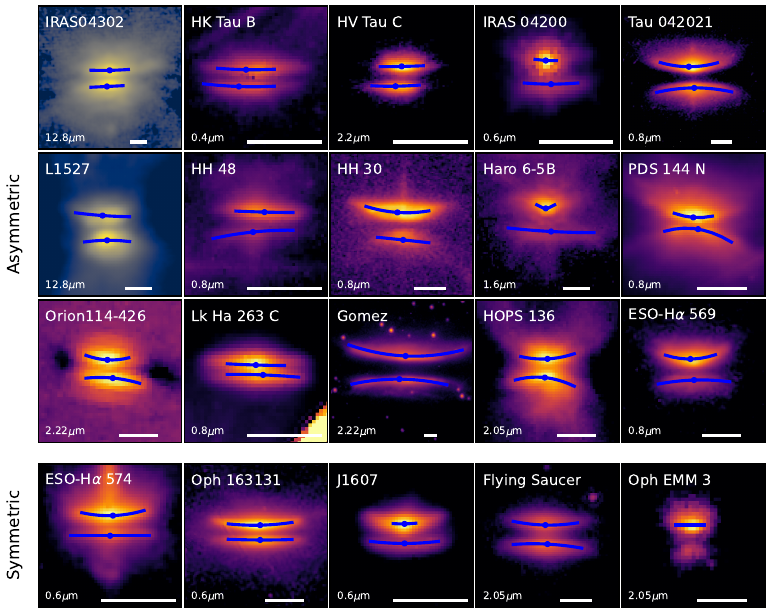}
    \caption{Gallery of edge-on disks with scattered light observations where we estimate the level of lateral asymmetry. Asymmetric disks have $|\delta_{spines}|\geqslant5\%$. The spines within 10\% of their respective maxima, used for the measurement of $\delta_{spines}$, are shown as blue lines. Their respective centers are marked by a point. A 1'' scale bar is also displayed in the bottom right of each panel. }
    \label{fig:symmetric-asymmetric}
\end{figure*}

\subsubsection{Other edge-on disks and general disk population}
\label{sec:comparisonotherdisks}
The models presented in Sect.~\ref{sec:model} showed that the impact of a tilted inner region is stronger at short wavelengths, where thermal emission is negligible. We found that for about half of the possible viewing orientations strong asymmetries are expected at 2$\mu$m, while a flux reversal can occur for about 1/3 of the possible orientations. These numbers may vary for different dust properties, inner disk sizes, and tilt geometry. Nevertheless, these predictions encourage us to analyse the diverse optical to near-infrared observations of edge-on protoplanetary disks, looking for any information on the occurrence of misaligned inner regions. 

We focus here on the identifications of lateral asymmetries in previous observations. Indeed, even though flux reversal is predicted to occur over a relatively large range of azimuthal orientation, it is less readily detectable. This is because observations over a wide range of wavelengths (ideally from the optical/near-infrared to the mid-infrared) are required to observe the brightest nebula switch at a particular viewing orientation. As highlighted in Sect.~\ref{sec:motivation} there are only 2/12 cases where brightest nebula switch was found between two wavelengths (IRAS40302, Flying Saucer).

Interestingly, a large number of scattered light observations of edge-on disks have revealed lateral asymmetries. To quantify the relative occurrence of lateral asymmetries, we analysed the list of 22 edge-on disks compiled by \citet[][their Table 5]{Angelo_2023}, which gathers known edge-on disks within 150~pc observed at optical or infrared wavelengths. We excluded 6 systems (IRAS04158+2805, HH390, ISO-Oph31, SSTc2d J162221.0-230402, DG Tau B, CB 26), either because the scattered light observations show a unipolar nebula or because no scattered light observations were previously published, and included 4 additional disks: PDS~144~N~\citep{Perrin_2006}, Orion114-426~\citep{McCaughrean_1998}, Gomez Hamburger~\citep{Bujarrabal_2008, Ruiz_1987}, and HOPS 136~\citep{Fischer_2014}. The final sample thus contains 20 disks, observed at scattered light wavelengths within the range of $0.4-21\mu$m.

For all disks, we performed a direct measurement of $\delta_{spines}$. Because of lower dynamic range, for 8 disks in the sample (HV Tau C,  IRAS04200, HH30, Haro6-5B, Orion114-426, J1607, Flying Saucer, Oph EMM 3), we measured the spine size within 20, 30 or 40\% of their respective peak, while for the rest of the sample we take the size within 10\% of the peaks. 
As in Sect.~\ref{sec:modelresults}, we identify disks with strong lateral asymmetries as those having $\delta_{spines} > 5\%$. 
We show the disk images, on which the spines are overplotted, in Fig.~\ref{fig:symmetric-asymmetric}, illustrating the prevalence of lateral asymmetries. 

We find that 15/20 disks show clear lateral asymmetries, while 5/20 of them do not. Lateral asymmetries at short wavelengths are a strong prediction of our radiative transfer models with a tilted inner region. Thus, the large number of edge-on disks observed with lateral asymmetries suggests that a significant fraction of protoplanetary disks may possess a tilted inner region. \\

In recent years, a number of protoplanetary disks observations have revealed tilted or warped inner regions. This was done mainly by detecting shadows in scattered light~\citep[e.g.,][]{Benisty_2018, Muro-Arena_2020} or finding gas kinematic signatures~\citep[e.g.,][]{Perez_2018}. Misalignement evidences are encountered both in young~\citep[e.g.,][]{Sakai_2019, Yamato_2023} and older protoplanetary disks~\citep[e.g.,][]{Debes_2017, Wolff_2016}, and the broad shadows moving over several year timescale in the TW~Hya disk~\citep{Debes_2023} suggests that they are potentially long lived. 

Statistical analysis of their occurrence is still limited but several studies of samples of few tens of disks have started to reveal a large fraction of misaligned systems. For example, \citet{Ansdell_2020} analysed a sample of 24 dipper objects, which are often assumed to host close to edge-on inner disks. They found that the outer disks are consistent with an isotropic distribution of inclinations, suggesting that most of these systems have misalignements. Besides, \citet{Bohn_2022} analysed the inclination of the inner and outer regions of a sample of 20 transition disks observed with VLT/GRAVITY and ALMA. They concluded that 6 disks were clearly misaligned, 5 were not, and were inconclusive about the 9 remaining systems. 
The analysis presented here showed that lateral asymmetries are frequent in edge-on protoplanetary disks (15/20). Given that tilted inner disks are also common around other systems, it is realistic that at least some of the asymmetry seen in edge-on disks are induced by a tilt. \\

Finally, we note that, even though some systems were observed multiple times and at different wavelengths, we limited our analysis to one observation only for each object (single epoch \& wavelength), meaning that time variability was not taken into account. 
Yet, \citet{Watson_2007_HH30} showed that, for example in HH30, the lateral asymmetry varies rapidly over a 10 years period. Other examples of variability are that the most recent 2023 observations of Tau042021 shows a symmetric disk~\citep{Duchene_2023}, as opposed to the asymmetric classification in Fig.~\ref{fig:symmetric-asymmetric}, while other systems are more stable (e.g., HK Tau B). In the case of HH30, the angle of observation of the system (azimuth) is not expected to change significantly over a 10 years timescale. If the lateral asymmetry is caused by a tilted inner disk scenario alone then it would imply that the inner disk have to precess very rapidly to explain the rapid changes in lateral differences, which might not be possible. 
Hence, more complete analysis and modeling would be needed to interpret the origin of the lateral asymmetry of each individual system. We discuss potential alternative explanations for the lateral asymmetries observed in edge-on disks in Sect.~\ref{sec:tilt_explanation}.

\subsubsection{Tilt origin and alternative explanations}
\label{sec:tilt_explanation}

Several mechanisms are thought to be able to lead to a warped disk. One of the first proposed mechanisms stated that misalignment of the rotation axis of the disk with the magnetic field direction can warp the innermost edge of the disk~\citep[AA Tau,][]{Bouvier_1999}. Alternatively, binaries misaligned with the disk or an inclined planet orbiting the central star, if massive enough, might also be able to generate misalignements between the inner and outer regions of the disk~\citep{Facchini_2013, Nealon_2018, Young_2023}. Other scenarios involve the anisotropic accretion or late capture of infalling material with a different angular momentum vector orientation than that of the disk~\citep{Dullemond_2019, Kuffmeier_2021}. Given the relatively young age of IRAS04302, it is not clear which scenario is most likely.

Alternative explanations for the asymmetries in the edge-on disks optical and near-infrared images have previously been proposed. For example, hot spots in the stellar surface can allow to reproduce the asymmetry in HST optical and near infrared images of HH30~\citep{Stapelfeldt_1999, Cotera_2001}. Such scenario would predict rapid asymmetry variations that may be detectable over few days timescales, as in the case of HH30. Hot spot might not lead to lateral asymmetries in the mid-infrared because their temperatures might be too hot, but a colder spot could. A test of this scenario with respect to the inner tilt scenario could include a variability study of the optical/near-infrared asymmetry (a tilted inner region would not rotate as fast as a stellar hot spot), and looking for the presence of a mid-infrared laterally asymmetry counterpart.  

Another possibility to generate lateral asymmetries includes the presence of asymmetric envelope or dust cloud material near the source but detailed models have not been performed to confirm these suggestions. Similarly, one could also imagine that the lateral asymmetries are due to an intrinsic morphological difference between the two sides of the disk (e.g., different flaring between the top/bottom or left/right sides). It is however not clear how such significant morphological differences could be long lived. For these scenarios, we would expect the lateral asymmetries to remain similar at all wavelengths, as opposed to predictions from the tilted models.

Finally, we note that the models with a tilted inner disk predict that lateral asymmetries are more pronounced and opposite between the near-infrared and the mid-infrared. Most disks were previously observed at optical to near-infrared wavelengths, and thus looking for asymmetries in these systems at 21$\mu$m can allow to test the misalignement hypothesis.

\subsection{Dark lane thickness and aspect ratio variation with evolutionary stage}
\label{sec:comparison_darklane}

The variation of the dark lane thickness of an edge-on disk is a hint towards its scale height and optical depth~\citep{Watson_2007}. One of the main goal of JWST GO program 2562 presented here (and upcoming cycle 2, JWST GO program 4290) is to use this measurement to obtain information on the level of dust settling and dust properties across different evolutionary stages. This could allow to better understand the timescale of dust accumulation of the midplane, and ultimately of planet formation. 

Here we aim to look for any trend for the variation of the dark lane thickness and of the disk aspect ratio at near- to mid-infrared wavelengths with evolutionary class. We focus on three edge-on disks, observed with JWST, in different evolutionary stages: the Class~I IRAS04302 studied in this work, L1527 (LDN 1527, Class 0/I), and Tau042021 (2MASS J04202144+2813491, Class~II). For L1527, we use archival 3.35$\mu$m, 4.44$\mu$m, 7.7$\mu$m, 12.8$\mu$m, and 18$\mu$m JWST observations and determine the disk morphology at the different wavelengths in this disk as in Sect.~\ref{sec:morphology} (see Appendix~\ref{app:l1527}). For Tau042021, we use the results obtained by \citet{Duchene_2023} with a similar methodology.

We consider both the dark lane thickness d$_{neb}$ and  the disk aspect ratio of the different observations. The latter is defined as the ratio of the dark lane thickness d$_{neb}$ and the averaged disk size at 7.7$\mu$m and 12.8$\mu$m. The disk size at one wavelength is taken as the smallest value of $R_{FW10\%}$ reported in Table~\ref{tab:darklane} for IRAS04302, Table~\ref{tab:darklane_l1527} for L1527, and the value of FW$10\%_{top}$ in Table 2 of \citet{Duchene_2023} for Tau042021. The final adopted disk sizes are 2.34''~\citep[237au, at 140pc;][]{vantHoff_2023}, 2.72''~\citep[438au, at 161pc;][]{Galli_2019}, and 2.63''~\citep[342au, at 130pc;][]{Galli_2019}, respectively for L1527, IRAS04302, and Tau042021. 

\begin{figure}
    \centering
    \includegraphics[width = 0.45\textwidth, trim=0cm 0cm 0cm 0cm,clip ]{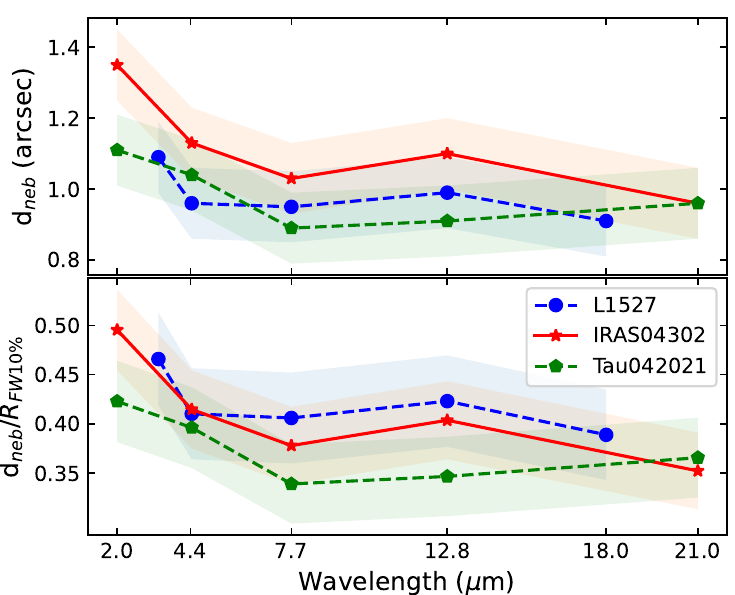}
    \caption{Dark lane thickness (top) or aspect ratio (bottom) as a function of wavelength in L1527 (Class~0/I), IRAS04302 (Class I), and Tau042021 (Class II). The uncertainties correspond to $\pm$0.1 for d$_{neb}$, propagated into the aspect ratio.}
    \label{fig:distances}
\end{figure}

\paragraph{Variation of the dark lane thickness} We show the variation of d$_{neb}$ with wavelength for these three objects in the top panel of  Fig.~\ref{fig:distances}. 
First, we find that although the disks have different radial extent, the separation of their nebulae appears to be very similar. Moreover, the dark lane thickness also does not vary significantly over this wavelength range (typically less than 10\%).

\citet{Duchene_2023} presented the first results of the series of observations of edge-on disks from JWST GO program 2562. They focused on the NIRCam and MIRI observations of Tau042021.  
In this system, they showed that the dark lane thickness does not vary significantly between 2$\mu$m and 21$\mu$m. 
By comparing the observations with predictions of a range of radiative transfer models, they concluded that such small variation indicates that grains have grown to at least 10$\mu$m and that such grains are not significantly settled. 
Although detailed radiative transfer would be required to take optical depth into account, the similar observational result obtained in  IRAS04302 and L1527 suggests that these disks have a similar configuration, with some level of grain growth in their upper layers.  
Consistent results of unsettled $\sim10\mu$m dust grains in the outer regions of protoplanetary disks have also been reported in previous studies of other edge-on disks~\citep[e.g.,][]{Pontoppidan_2007, Sturm_2023}.

At other locations than in the disk midplane, \citet{Franceschi_2023} showed that the collisional timescale in protoplanetary disks is typically larger than disk lifetimes, preventing efficient dust growth in the upper layers of the disk. Finding intermediate grains in the upper layers of disks could thus suggest that either turbulence in these systems is  sufficiently strong to replenish the upper layers with larger grains from the midplane~\citep[e.g.,][]{Tazaki_2021, Sturm_2023}, or that dust particles are able to grow there through processes other than turbulent collision~\citep[e.g., coagulation driven by settling;][]{Franceschi_2023}.

\paragraph{Variation of the aspect ratio} The bottom panel of Fig.~\ref{fig:distances} shows the variation of the disk aspect ratio with wavelength, which allows to reduce potential dependencies of the apparent height with disk radius. We find that the disk aspect ratio at JWST wavelengths tentatively decreases with evolutionary class. At all wavelengths, the young L1527 appears relatively thicker than IRAS04302, which is also thicker than the older Tau042021, although uncertainties are large.

The apparent height of the dust grains observed with JWST is linked to the total disk dust mass and the gas scale height, which itself depends on the disk temperature and stellar mass. 
Tau042021 and L1527 have a similar stellar mass, but L1527 has both a higher dust mass ($M_{dust}>41$M$_\oplus$ vs $M_{dust}>25.5$M$_\oplus$; \citealt{vantHoff_2023}, \citealt{Villenave_2020}) and possibly higher disk temperature than Tau042021 \citep[CO freeze out at $R>350$au vs $R>100$au,][]{vantHoff_2023, Duchene_2023}, which could explain its higher aspect ratio. On the other hand, the difference in aspect ratio between Tau042021 and IRAS04302 is possibly driven by the larger dust mass in IRAS04302 ($M_{dust}>54.8$M$_\oplus$ vs $M_{dust}>25.5$M$_\oplus$; \citealt{Villenave_2020}).  
Altogether the tentative decrease in disk aspect ratio with evolutionary class may be linked to changes in the disk temperature structure, dust mass, or dust properties with age~\citep[see e.g.,][]{vant_Hoff_2020}. 

\section{Conclusions}
\label{sec:conclusion}

In this study, we have presented new JWST observations of the Class~I IRAS04302+2247, obtained with both NIRCam and MIRI imagers. The source is detected in scattered light at all wavelengths. At 2$\mu$m and 4.44$\mu$m the disk appears surrounded by a prominent envelope, which gradually disappears at 7.7$\mu$m, leaving clear disk images at 12.8$\mu$m and 21$\mu$m.
We compared 25.5 year older HST observations with the new JWST 2.0$\mu$m image and identify a scattered light blob feature which is moving slowly, directly away from the presumed star location. The analysis of its proper motion indicates that it is likely not a background source, and its slow velocity suggests that it corresponds to material entrained by the low-velocity outflow. 

We characterized the morphology of IRAS04302 and two other disks of different evolutionary classes (L1527 and Tau042021) and found no significant variation of dark lane thickness with wavelength (typically within $\sim 10$\%). This is consistent with the presence of relatively large grains (several microns) unsettled in the upper layers of the disk \citep{Duchene_2023}. In addition, the aspect ratio of the disks seems to decrease with evolutionary stage, which could trace evolution of the disk temperature structure, disk mass, and/or grain properties with age.

The characterization of IRAS04302 also revealed that the disk appears asymmetric, with its western nebula more extended to the south than its eastern counterpart. Finally, we find that the brightest nebula switches side between 12.8$\mu$m and 21$\mu$m. 

We performed radiative transfer models introducing a tilted inner region to explore this particular aspect of the observations. We find that a tilted inner region can lead to a switch in the brightest nebula for about 1/3 of the possible viewing orientation,  and that such a model is able to reproduce the MIRI observations of IRAS04302. We discuss the viewing orientation of the system based on the modeling. 

In addition, for some orientations, the model with a tilted inner disk predicts strong lateral asymmetries in edge-on disks. The intensity and orientation of such asymmetry depend on the wavelength, and are opposite between the near- and the mid-infrared. At short wavelengths (2$\mu$m) about half of all possible viewing orientations are susceptible to create strong lateral asymmetries. 
We identify 15/20 edge-on disks in the literature with such lateral asymmetries. 
While stellar spots may also create lateral asymmetries, the frequency of asymmetries suggests that a large fraction of protoplanetary disks might possess misaligned inner regions, consistent with previous observational results. 
Variability studies of edge-on disks could allow to test the dominant scenario. Indeed, stellar spots would lead to asymmetric variations over several days timescale, while a tilted inner disk might not lead to variation over year to decade timescale.  
Observations at longer wavelengths (21$\mu$m), such as enabled by the upcoming JWST GO program 4290, could also allow to confirm the presence of a tilted inner region in these asymmetric disks. Indeed, the model predicts that asymmetries would be opposite at 21$\mu$m than at shorter wavelengths.

\medskip
\emph{Acknowledgements:}
We thank Howard Bushouse and Ken MacDonald at STScI for fixing the pixel fitting issues in the JWST pipeline and allowing to recover the saturated pixels in the F444W image. We are grateful to the W. M. Keck Institute for Space Studies as the hosts for our week-long team meeting in May 2023.  
This project has received funding from the European Research Council (ERC) under the European Union's Horizon Europe research and innovation program (grant agreement No. 101053020, project Dust2Planets, PI F. M\'enard).
MV, KRS, GD, and SGW acknowledge funding support from JWST GO program \#2562 provided by NASA through a grant from the Space Telescope Science Institute, which is operated by the Association of Universities for Research in Astronomy, Incorporated, under NASA contract NAS5-26555.
The research of M.V. was supported by an appointment to the NASA Postdoctoral Program at the NASA Jet Propulsion Laboratory, administered by Oak Ridge Associated Universities, under contract with NASA. RT acknowledges financial support from a CNES postdoctoral fellowship.  
The IRAS04302+2247 data presented in this paper were obtained from the Mikulski Archive for Space Telescopes (MAST) at the Space Telescope Science Institute. The specific observations analyzed can be accessed via \dataset[10.17909/kh9r-ym63]{https://doi.org/10.17909/kh9r-ym63}.


\vspace{5mm}
\facilities{JWST, HST}
\software{\texttt{mcfost}~\citep{Pinte_2006, Pinte_2009}, \texttt{Matplotlib}~\citep{Hunter_2007}, \texttt{Numpy}~\citep{Harris_2020}.}

\appendix

\section{JWST observations of L1527}
\label{app:l1527}

In Sect.~\ref{sec:comparison_darklane}, we use JWST NIRCam and MIRI data of L1527 observed as part of DD program 2739 (PI: K. Pontoppidan). We consider observations at 5 wavelengths, taken with filters F335M (3.35$\mu$m), F444W (4.44$\mu$m), F770W (7.7$\mu$m), F1280W (12.8$\mu$m) and F1800W (18$\mu$m). The MIRI data were observed successively on September 1$^{st}$, 2022, with about 45 minutes exposures per filter. On the other hand, NIRCam observed the source on September 8, 2023 for 1460s.

We obtained the phase 2 pipeline calibrated data from the MAST archive, and we re-run phase 3, using the JWST pipeline version 1.9.5. Similarly to the reduction of IRAS04302 presented in Sect.~\ref{sec:reduction}, we set \texttt{subtract = True} in the \texttt{skymatch} step to subtract the background. The average background levels were 0.26, 0.53, 23.0, 85.08, and 170.23 MJy/sr at F335W, F444W, F770W, F1280W, and F1800W, respectively. 

\begin{figure*}
    \centering
    \includegraphics[width = 1\textwidth, trim=0.5cm 0.1cm 0cm 0.5cm,clip ]{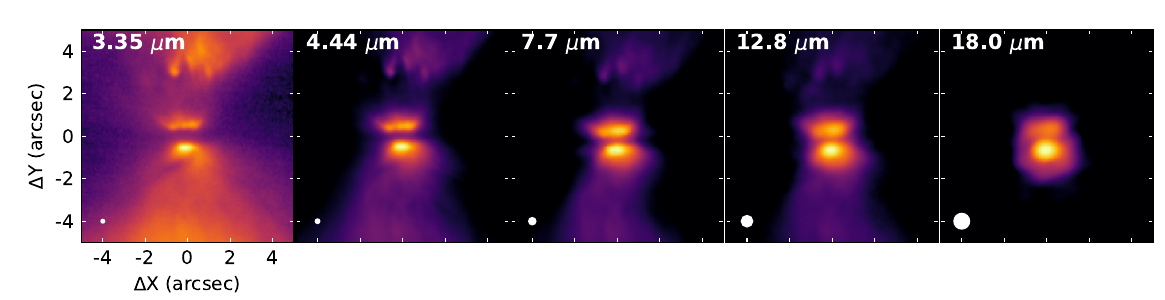}
    \caption{Gallery of the JWST observations of L1527. All images are shown with a log stretch. The bottom left ellipse in each panel indicates the corresponding angluar resolution of the observations.}
    \label{fig:l1527}
\end{figure*}

\begin{table*}
    \caption{L1527 morphological properties.}
    \centering
    \begin{tabular}{lcccccccccccc}
    \hline\hline
    $\lambda$ & $F_{5^{\prime\prime}\times5^{\prime\prime}} $& $d_{neb}$ & FR &  $R_{FWHM}$ &$R_{FWHM}$ & $R_{FW10\%}$ & $R_{FW10\%}$ & $\delta_{spines}$\\
    ($\mu$m)& (mJy) & (\asec) & T/B& T & B &T & B & T-B (\%)\\
    \hline
    3.35 & $0.3 \pm 0.03$ & $1.09 \pm 0.10$ & $0.41 \pm 0.05$ & $1.18 \pm 0.10 $ & $1.81 \pm 0.17$ & $-$ & $-$ & $-$\\
    4.44 & $8 \pm 1$ & $0.96 \pm 0.10$ & $0.50\pm 0.04$ & $0.92\pm 0.10$ & $1.63\pm 0.10$ & $2.18 \pm 0.10$ & $2.63 \pm 0.10$ & -9\\
      7.7 & $ 13 \pm 1$ & $0.95\pm 0.10$ & $0.64 \pm 0.01$ &  $1.18 \pm 0.10$ & $1.42 \pm 0.10$ & $2.22 \pm 0.10$ & $2.45 \pm 0.10$ & -13\\
     12.8 & $10 \pm1$ &$0.99 \pm 0.10$ & $0.51\pm 0.03$  & $1.15 \pm 0.10$ & $1.37 \pm 0.10$ &  $2.46 \pm 0.11$ & $2.67 \pm 0.13$ & -12\\
     18.0& $72 \pm 7$ & $0.91\pm 0.10$ & $0.29 \pm 0.10$  & $1.24\pm 0.10$ & $-$ & $-$ & $-$  & $-$ \\
    \hline
    \end{tabular}
    \tablecomments{Same parameters as in Table~\ref{tab:darklane}. A negative value of $\delta_{spines}$ indicates that the bottom nebula is centered further West than the top nebula. The error bars of $d_{neb}$, $R_{FWHM}$, and $R_{FW10\%}$ correspond to the statistical error for the different spine averaging or 0.1\asec, whichever is greater.  "T" stands for the top nebula, and "B" for bottom nebula in Fig.~\ref{fig:l1527}. }
    \label{tab:darklane_l1527}
\end{table*}

The final images are shown in Fig.~\ref{fig:l1527}. The disk is vertically and radially resolved up to 18$\mu$m. Except for the observations at 18$\mu$m all other images show extended scattered light along the minor axis direction. This feature is connected to the spectacular outflow cavity revealed by the NIRCam observations\footnote{https://www.nasa.gov/feature/goddard/2022/nasa-s-webb-catches-fiery-hourglass-as-new-star-forms}. Moreover, we see that the disk is not laterally symmetric. Specifically, the top nebula appears to extend towards the left while the bottom nebula is wider to the right. This asymmetry may be due to a tilted inner region as we argue in Sect.~\ref{sec:comparisonotherdisks}. Interestingly, previous millimeter continuum and large scale $^{12}$CO observations also suggested the presence of a misaligned inner region in this system~\citep{Sakai_2019, vantHoff_2023}.

We measure the flux within an squared aperture  of 5\asec. Then, we apply the same methodology as in Sect.~\ref{sec:morphology} to determine the morphological features of the disk, namely dark lane thickness ($d_{neb}$), the peak flux ratio of the bottom over the top nebula (FR), the radial $R_{FWHM}$ and $R_{FW10\%}$ of the nebulae, and $\delta_{spines}$ to quantify the lateral asymmetry. The results are summarized in Table~\ref{tab:darklane_l1527}.

\bibliography{biblio}{}
\bibliographystyle{aasjournal}

\end{document}